\documentclass[a4paper,11pt]{article}
\pdfoutput=1

\usepackage{jheppub}

\usepackage[abs]{overpic}

\usepackage{warpcol}

\usepackage{lineno}
\usepackage{graphicx}
\usepackage{dcolumn}
\usepackage{bm}
\usepackage{rotating}
\usepackage{epstopdf}
\usepackage{color}
\usepackage{verbatim} 
\usepackage{multirow}
\usepackage[abs]{overpic}
\usepackage{amsmath}
\usepackage{mathrsfs}
\usepackage{amssymb}
\usepackage{subfigure}
\usepackage{xspace}
\usepackage{float}
\usepackage{hyperref}
\hypersetup{colorlinks=true, linkcolor=blue, anchorcolor=blue, citecolor=blue}

\newcommand{\PreserveBackslash}[1]{\let\temp=\\#1\let\\=\temp}
\newcolumntype{C}[1]{>{\PreserveBackslash\centering}p{#1}}
\newcolumntype{R}[1]{>{\PreserveBackslash\raggedleft}p{#1}}
\newcolumntype{L}[1]{>{\PreserveBackslash\raggedright}p{#1}}

\newcommand{\pp}{\pi^+\pi^-}

\newcommand{\EE}{e^+e^-}

\newcommand{\GG}{\gamma\gamma}

\newcommand{\jpsi}{J/\psi}
\newcommand{\piz}{\pi^{0}}
\newcommand{\ppp}{\pi^{+}\pi^{-}\pi^{0}}
\newcommand{\too}{\rightarrow}
\usepackage[T1]{fontenc}

\RequirePackage{lineno}
\usepackage{epstopdf}

\title{\boldmath Improved measurement of the branching fraction of $h_{c}\too\gamma\eta^\prime/\eta$ and search for $h_{c}\too\gamma\piz$}

\collaborationImg{\includegraphics[width=.12\textwidth,origin=c,angle=90]{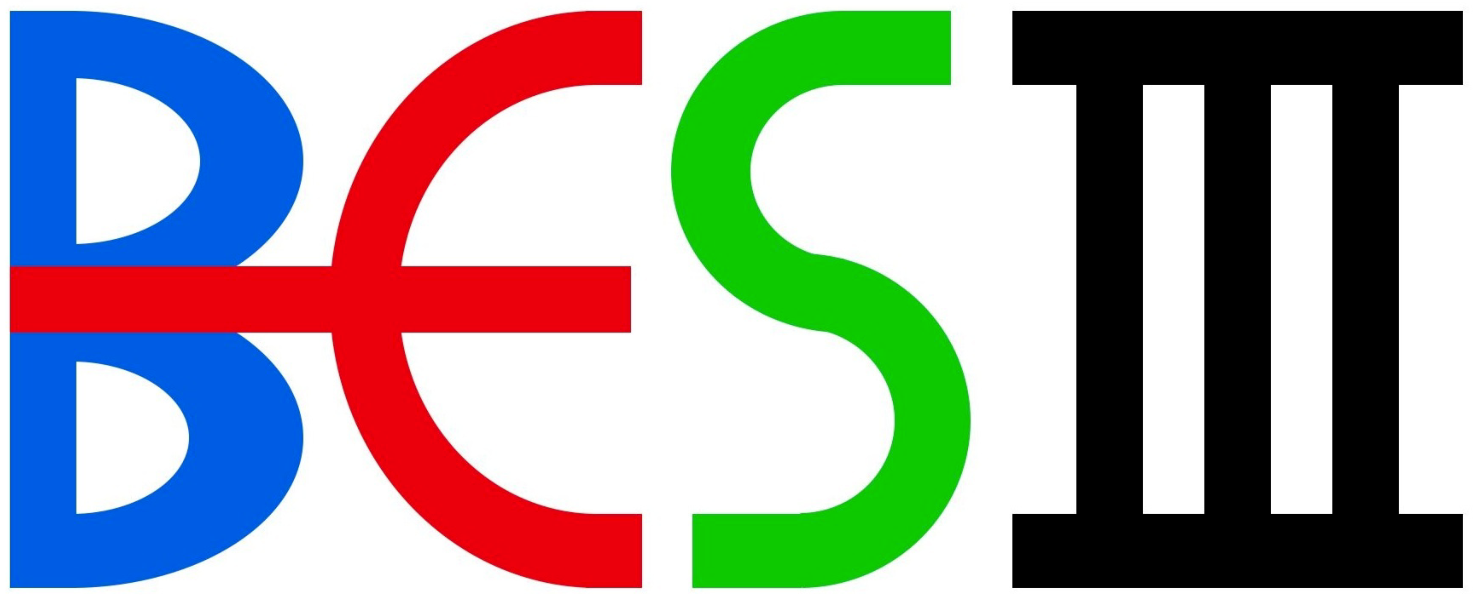}}

\collaboration{BESIII Collaboration}


\emailAdd{besiii-publications@ihep.ac.cn}

\arxivnumber{2405.11585}

\abstract{
The processes $h_c\to\gamma P(P = \eta^\prime,~\eta,~\piz)$ are studied with a sample of $(27.12\pm0.14)\times10^{8}$ $\psi(3686)$ events collected by the BESIII detector at the BEPCII collider. The decay $h_{c}\too\gamma\eta$ is observed for the first time with the significance of $9.0\,\sigma$, and the branching fraction is determined to be $(3.77\pm0.55\pm0.13\pm0.26)\times10^{-4}$, while $\mathscr{B}(h_{c}\too\gamma\eta^\prime)$ is measured to be $(1.40\pm0.11\pm0.04\pm0.10)\times10^{-3}$, where the first uncertainties are statistical, the second systematic, and the third from the branching fraction of $\psi(3686)\too\pi^{0}h_c$. The combination of these results allows for a precise determination of $R_{h_c}=\frac{\mathscr{B}(h_c\too\gamma\eta)}{\mathscr{B}(h_c\too\gamma\eta^\prime)}$, which is calculated to be $(27.0\pm4.4\pm1.0)\%$. The results are valuable for gaining a deeper understanding of $\eta-\eta^\prime$ mixing, and its manifestation within quantum chromodynamics.   No significant signal is found for the decay $h_c\too\gamma\pi^{0}$, and an upper limit is placed on its branching fraction of  $\mathscr{B}(h_c\too\gamma\pi^{0})<5.0\times10^{-5}$, at the 90\% confidence level.}

\begin{document}
\maketitle
\flushbottom

\section{Introduction}

The study of $\eta-\eta^\prime$ mixing is a poweful laboratory for testing the $U(1)_A$ anomaly~\cite{u1a} and $\rm SU(3)$ breaking~\cite{u3f} in Quantum Chromodynamics (QCD), and also probes the gluonic content of the $\eta^{(\prime)}$ mesons.
Assuming that the light quarks are massless, the $\eta^\prime$ meson is a pure flavor singlet. Due to  flavor-symmetry breaking, there is mixing among the neutral mesons. Without the $U(1)_A$ anomaly, the two iso-singlet mass eigenstates in the pseudoscalar sector would consist of $u\bar{u} + d\bar{d}$ and $s\bar{s}$, respectively. With the $U(1)_A$ anomaly present, these states mix, resulting in the nearly flavor octet or singlet combinations that correspond to the physical $\eta$ and $\eta^\prime$ mesons, respectively.
This mixing has attracted a great deal of experimental and theoretical interest. From the experimental side, the $\eta-\eta^\prime$ mixing can be studied through the radiative decays of charmonia to $\eta$ and $\eta^\prime$. Recent measurements have mostly focused on the decay of S-wave charmonium states, such as $J/\psi(\psi(3686))\too\gamma\eta$ and $\gamma\eta^\prime$~\cite{jpsigam, psipgam}. From the theoretical side, most of the predictions concerned with this topic regarding branching fractions are compatible with experimental data~\cite{brtheo}. However, the $\eta-\eta^\prime$ mixing angle extracted from $J/\psi\too\gamma\eta$ and $\gamma\eta^\prime$ decays, are not consistent among the  different theoretical approaches~\cite{diffangle1, diffangle2, diffangle3, diffangle4}. These discrepancies indicate that our understanding of $\eta-\eta^\prime$ mixing is not complete, and that further experimental and theoretical investigations are necessary.

The $P$-wave singlet charmonium state, $h_{c}$, possesses spin-parity quantum numbers that differ from those of the $J/\psi$. This distinction leads to different characteristics and behavior in its radiative decays and other processes, compared to those of the vector-meson state. The gluonic contributions of $\eta^{(\prime)}$ are significantly suppressed in the radiative decays of the $J/\psi$~\cite{diffangle2}. However, in the radiative decays of $h_c$, this suppression effect is absent. Therefore, the radiative decays of the $h_c$ provide additional insights and information for understanding of the gluonic contribution of the $\eta^{(\prime)}$ mesons.
The $h_c$ radiative decays, $h_c\too\gamma\eta^\prime$ and $h_c\too\gamma\eta$, have been investigated at BESIII through the decay $\psi(3686)\too\pi^{0}h_c$. The statistical significances obtained for $h_{c}\too\gamma\eta^\prime$ and $h_{c}\too\gamma\eta$ are $8.4\,\sigma$ and $4.0\,\sigma$, respectively~\cite{hcgamX}. These results have stimulated extensive theoretical discussion~\cite{hcb, hcmlp, pqcd, mixpqcd}.
The $\eta-\eta^\prime$ mixing angle, associated with the ratio $R_{h_c}=\mathscr{B}(h_c\too\gamma\eta)/\mathscr{B}(h_c\too\gamma\eta^\prime)$, is determined to be $\phi=33.8^{\rm o}\pm2.5^{\rm o}$ through a perturbative QCD calculation~\cite{mixpqcd}. This value is compatible with recent lattice QCD calculations~\cite{diffangle4, diffangle5}, but deviates from other calculations~\cite{diffangle1, diffangle3}. Since the uncertainty of the mixing angle $\phi$ is dominated by $R_{h_c}$, a more precise measurement of this ratio is essential.

Knowledge of the properties of the $h_c$ is poor relative to that of  other hidden charmonium states, e.g. the $J/\psi$. According to the Particle Data Group (PDG)~\cite{pdg}, excluding the E1 decay $h_c\too\gamma\eta_c$, which has a branching fraction of $57\pm5$\%, the sum of the branching fractions of known $h_c$ decay modes is only about 3\%. Studies of the radiative decays $h_c\to\gamma P (P=\eta^\prime,~\eta,~\piz)$ are particularly desirable as these will provide more direct information on non-perturbative effects in heavy quarkonia~\cite{hcb}. Previous measurements of the branching fractions of $h_c\to\gamma\eta^\prime(\eta)$ have limited precision~\cite{hcgamX}, as listed in TABLE~\ref{tab:result}, and the decay $h_c\to\gamma\piz$ remains unobserved.

In this paper,  we present the improved measurements of the branching fractions of $h_c\too\gamma\eta^\prime(\eta)$ and a search for $h_c\too\gamma\piz$ via $\psi(3686)\too\piz h_{c}$, based on $(27.12\pm0.14)\times10^{8}$ $\psi(3686)$ events~\cite{0912data, 21data} collected by the BESIII detector~\cite{besiii}, which is about 6 times more than previously collected~\cite{0912data}.

\section{BESIII detector and Monte Carlo simulation}

The BESIII detector~\cite{besiii} records symmetric $e^+e^-$ collisions provided by the BEPCII storage ring~\cite{bepcii} in the center-of-mass energy range from 2.0 to 4.95~GeV, with a peak luminosity of $1 \times 10^{33}\;\text{cm}^{-2}\text{s}^{-1}$ achieved at $\sqrt{s} = 3.77\;\text{GeV}$. BESIII has collected large data samples in this energy region~\cite{highE}. The cylindrical core of the BESIII detector covers 93\% of the full solid angle and consists of a helium-based multilayer drift chamber~(MDC), a plastic scintillator time-of-flight system~(TOF), and a CsI(Tl) electromagnetic calorimeter~(EMC), which are all enclosed in a superconducting solenoidal magnet providing a 1.0~T magnetic field. The solenoid is supported by an octagonal flux-return yoke with resistive plate counter muon identification modules interleaved with steel. The charged-particle momentum resolution at $1~{\rm GeV}/c$ is $0.5\%$, and the ${\rm d}E/{\rm d}x$ resolution is $6\%$ for electrons from Bhabha scattering. The EMC measures photon energies with a resolution of $2.5\%$ ($5\%$) at $1$~GeV in the barrel (end-cap) region. The time resolution in the TOF barrel region is 68~ps, while that in the end-cap region was 110~ps. The end cap TOF system was upgraded in 2015 using multigap resistive plate chamber technology, providing a time resolution of 60~ps~\cite{etof}.

Simulated data samples produced with a {\sc geant4}-based~\cite{geant4} Monte Carlo (MC) package, which includes the geometric description of the BESIII detector and the detector response, are used to determine the detection efficiency and to estimate the background contributions. The simulation includes the beam-energy spread and initial-state radiation (ISR) in $e^+e^-$ annihilations modeled with the generator {\sc kkmc}~\cite{KKMC}. The inclusive MC sample includes the production of the $\psi(3686)$ resonance, the ISR production of the $\jpsi$, and the continuum processes incorporated in {\sc kkmc}~\cite{KKMC}. All particle decays are modeled with {\sc evtgen}~\cite{ref:evtgen} using branching fractions either taken from PDG~\cite{pdg}, when available, or otherwise estimated with {\sc lundcharm}~\cite{ref:lundcharm}. Final-state radiation from charged final state particles is incorporated using the {\sc photos} package~\cite{photos}.

The signal MC samples for the decay $\psi(3686)\too\piz h_{c}$, $h_{c}\too\gamma\eta^\prime(\eta,~\piz)$ are generated by {\sc evtgen}~\cite{ref:evtgen}. The decay $\eta^\prime\too\gamma\pi^{+}\pi^{-}$ is simulated using the DIY model~\cite{etapgam2pi}, which takes $\rho-\omega$ interference and the box anomaly into account. The decay $\eta^\prime\too\eta\pi^{+}\pi^{-}$ is generated 
using the Dalitz distribution of this decay~\cite{etapeta2pi}
with the subsequent decay of $\eta\too\gamma\gamma$ produced by a phase-space model. The $\eta$ directly from $h_{c}$ decay is reconstructed via both $\eta\too\gamma\gamma$ and $\eta\too\pi^{+}\pi^{-}\piz$, in which the latter is generated using the Dalitz distribution as measured in Ref.~\cite{eta3pi}. The $\piz\too\gamma\gamma$ decay is also generated using the phase-space model.

\section{Event section and background analysis}

Charged tracks detected in the MDC are required to be within a polar-angle ($\theta$) range of $(|\!\cos\theta|<0.93)$, where $\theta$ is defined with respect to the $z$ axis, which is the symmetry axis of the MDC. The distance of the closest approach to the interaction point (IP) must be less than 10~cm along the $z$-axis, $V_{z}$, and less than 1~cm in the transverse plane, $V_{xy}$. Photon candidates are identified using showers in the EMC. The deposited energy of each shower must be more than 25 MeV in the barrel region ($|\!\cos\theta|<0.80$) and more than 50~MeV in the end-cap region ($0.86<|\!\cos\theta|<0.92$). To exclude showers that originate from charged tracks, the angle subtended by the EMC shower and the position of the closest charged track at the EMC must be greater than 10~degrees as measured from the IP. To suppress electronic noise and showers unrelated to the event, the difference between the EMC time and the event start time is required to be within [0, 700] ns. If there is no charged particle in the final state, e.g. $h_{c}\too\gamma\eta$, $\eta\too\gamma\gamma$, the timing information from EMC cluster is required to be within $|T-T_{\rm max}| \leq 500$ ns, where $T$ is the time associated with the EMC cluster for each photon candidate, and $T_{\rm max}$ is the time of the EMC cluster associated with the most energetic photon.

For the decay $\psi(3686)\too\piz h_{c}$ with $h_{c}\too\gamma\eta^\prime(\eta^\prime\too\pi^{+}\pi^{-}\eta)$ or $h_{c}\too\gamma\eta(\eta\too\pi^{+}\pi^{-}\piz)$, the final states for both processes consist of a $\pi^{+}\pi^{-}$ pair and five photons. Candidate events are required to have two charged tracks with zero net charge and at least five photons. In the event pre-selection, the $\piz$ mass window is set to be $[0.10,~0.16]\,{\rm GeV}/c^{2}$, and the $\eta$ from the $\eta^\prime$ decay is reconstructed by the $\GG$ final state within a mass window of $[0.45,~0.59]\,{\rm GeV}/c^{2}$. To reduce contamination and improve the mass resolution, a six-constraint (6C) kinematic fit is performed, which comprises a four-constraint (4C) fit with  two additional mass constraints.  The 4C fit requires the total four-momentum of charged tracks and photon candidates to be equal to the initial $\psi(3686)$ momentum.  In addition, the invariant mass of the $\GG$ pair in the decay of $\psi(3686) \rightarrow \pi^{0} h_{c}$ is constrained to the known $\pi^{0}$ mass~\cite{pdg}, and the invariant mass of  the $\GG$ pair from the $h_c$ decay is constrained to the known $\pi^0$ or $\eta$ mass~\cite{pdg}, according to the decay chain being analysed.

For the $\psi(3686)\too\piz h_{c}$, $h_{c}\too\gamma\eta(\eta\too\pi^{+}\pi^{-}\piz)$ decay, there are two $\piz$ candidates in the final state. The higher energy $\piz$ is taken as the one decaying from the $\psi(3686)$. In the case of multiple candidates in an event, the one with the smallest $\chi^{2}_{\text{6C}}$ value is selected. The $\chi^{2}_{\text{6C}}$ value of $h_{c}\too\gamma\eta^\prime(\eta^\prime\too\pi^{+}\pi^{-}\eta)$ or $h_{c}\too\gamma\eta(\eta\too\pi^{+}\pi^{-}\piz)$ is required to be less than 70 or 40, respectively. To suppress background events with $\gamma\pi^{+}\pi^{-}\eta\piz$ final states, such as those arising from the decays  $\psi(3686)\too\gamma\chi_{c2},~\chi_{c2}\too\eta\eta$, one $\eta$ decays to $\gamma\gamma$, another decays to $\pi^{+}\pi^{-}\pi^0$, an additional requirement of $\chi^{2}_{\text{6C}}{(\gamma\pi^{+}\pi^{-}\pi^0\pi^{0})} < \chi^{2}_{\text{6C}}(\gamma\pi^{+}\pi^{-}\eta\pi^0)$ is imposed for the signal candidates of $h_{c}\too\gamma\eta(\eta\too\pi^{+}\pi^{-}\piz)$.
Here $\chi^{2}_{\text{6C}}{(\gamma\pi^{+}\pi^{-}\pi^0\pi^{0})}$ is the $\chi^{2}$ value of 6C kinematic fit under the hypothesis of $\gamma\pi^{+}\pi^{-}\piz\piz$, while $\chi^{2}_{\text{6C}}(\gamma\pi^{+}\pi^{-}\eta\pi^0)$ is the $\chi^{2}$ value of 6C kinematic fit under the hypothesis of $\gamma\pi^{+}\pi^{-}\eta\piz$.

For the decay of $h_{c}\too\gamma\eta^\prime(\eta^\prime\too\gamma\pi^{+}\pi^{-})$, candidate events are required to have two charged tracks with zero net charge and at least four photons. A 5C kinematic fit, combining a 4C fit with an additional 1C constraint for the $\piz$ mass, is applied to the decay chain and used to loop over all possible combinations of photons. The combination with the smallest $\chi^{2}_{\text{5C}}$ value is selected and is required to be less than 40. The highest-energy radiative-photon candidate is assigned as coming from the $h_c$ decay, and the lowest-energy candidate to the $\eta^\prime$ decay.

In the analysis of the fully neutral final states, $\psi(3686)\too\piz h_{c}$, where $h_{c}$ decays into either $\gamma\eta$($\eta\too\gamma\gamma$) or $\gamma\piz$($\piz\too\gamma\gamma$), at least five photons are required. For the $h_{c}\too\gamma\eta$($\eta\too\gamma\gamma$) decay, a 5C kinematic fit, combining a 4C fit with an additional 1C constraint for the $\piz$ mass, is used to select the best candidate with the minimum $\chi^{2}_{\text{5C}}$ value by looping over all possible photon combinations. Additionally, the $\chi^{2}_{\text{5C}}$ value is required to be less than 40.
The highest energy photon among the three candidates remaining after excluding those two used to reconstruct the $\piz$ decay, is taken to  be the radiative photon from the $h_{c}$ decay.
When selecting $h_{c}\too\gamma\piz$($\piz\too\gamma\gamma$) decays, a 6C kinematic fit is performed, which includes a 4C fit and additional 2C constraints, requiring that the invariant mass of each $\gamma\gamma$ pair equals that of the known $\piz$ mass. We loop over all possible combinations of photons and select those that give the minimum $\chi^{2}_{\text{6C}}$ value.  The higher energy of the two $\pi^0$ candidates is selected as the one originating from the $h_{c}$ decay. The $\chi^{2}_{\text{6C}}$ value is required to be less than 20. The $\chi^{2}$ requirements for the $h_{c}\too\gamma\eta^\prime(\eta)$ selection have been optimized by maximizing the $S/\sqrt{S+B}$ figure of merit, where $S$($B$) is the number of signal (background) events in the signal region from the normalized signal (inclusive) MC sample. As there is no significant signal for the decay $h_{c}\too\gamma\piz$, the figure of merit in this case is defined as $\epsilon/(3/2+\sqrt{B})$~\cite{fom}, where $\epsilon$ is the signal efficiency.

The remaining dominant background for the $h_{c}\too\gamma\eta$, $\eta\too\gamma\gamma$ decay arises from a fake photon combined with a real photon, thereby forming a fake $\eta$ signal. This contamination is suppressed by requiring the ratio $R_{E} =\frac{|E_{\gamma1}-E_{\gamma2}|}{P_{\eta}}<0.94$, where $E_{\gamma1}$ and $E_{\gamma2}$ are the energies of the two photons from the $\eta$ candidate decay, and $P_{\eta}$ is the momentum of the $\eta$ candidate in the laboratory frame. For the $h_{c}\too\gamma\piz$ decay, there are five photons in the signal final states. Four of them come from the two $\pi^{0}$, and one bachelor photon is from the $h_{c}$ radiative decay. To suppress this bachelor photon combining with other photons to form fake $\pi^{0}$ signal, we require $M(\gamma\gamma_{{\rm low}})>0.16$ GeV/$c^{2}$, where $\gamma$ is the radiative photon, and $\gamma_{{\rm low}}$ is the low-energy photon from $\pi^{0}$ in the decay of $\psi(3686)\too\pi^{0}h_{c}$. Another source of background comes from the process $\psi(3686)/\EE\too\gamma X$ with $X\too\pi^{0}_{1}\pi^0_{2}$, where $X$ represents possible intermediate states, $\pi^{0}_{1}$ and $\pi^{0}_{2}$ are the $\pi^{0}$ with low energy and high energy, respectively. This contamination is suppressed by requiring the cosine of the angle between $\piz_{1}$ and $\piz_{2}$, denoted as $\cos\theta_{\pi^{0}_{1}\pi^{0}_{2}}$, to be larger than $-0.5$.

\section{Fit and numerical results}

\begin{figure}[htbp]
\begin{center}
\hspace*{-0.3cm}
\begin{overpic}[width=0.9\textwidth]{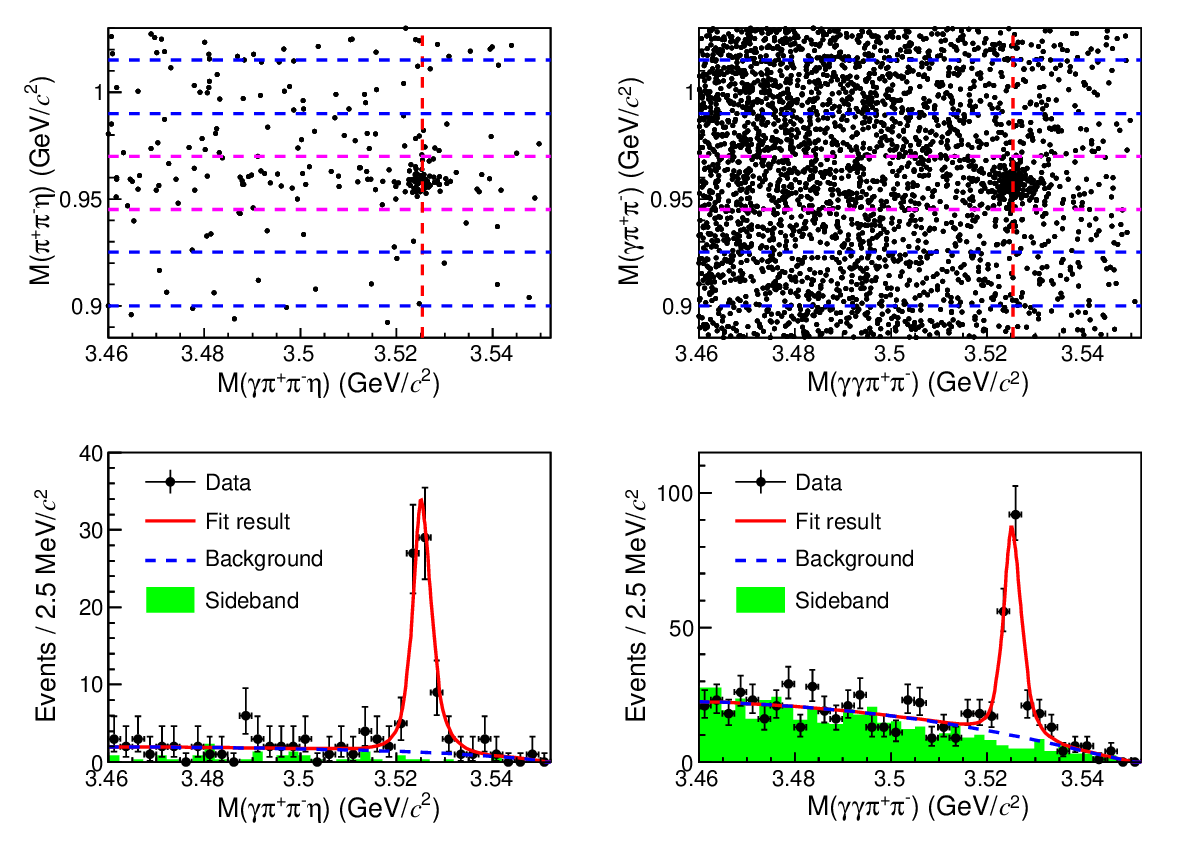}
\put(150,250){\footnotesize{\textbf{(a)}}}
\put(350,250){\footnotesize{\textbf{(b)}}}
\put(150,110){\footnotesize{\textbf{(c)}}}
\put(350,110){\footnotesize{\textbf{(d)}}}
\end{overpic}
\caption{Distributions of (a) $M(\gamma\pi^{+}\pi^{-}\eta)$ versus $M(\pi^{+}\pi^{-}\eta)$ of the candidates for $h_{c}\too\gamma\eta^\prime$ with $\eta^\prime\too\pi^{+}\pi^{-}\eta$, and (b) $M(\gamma\gamma\pi^{+}\pi^{-})$ versus $M(\gamma\pi^{+}\pi^{-})$ of the candidates for $h_{c}\too\gamma\eta^\prime$ with $\eta^\prime\too\gamma\pi^{+}\pi^{-}$. The horizontal pink (blue) dashed lines mark the signal (sideband) region(s) of $\eta^\prime$ and the vertical red dashed lines mark the known $h_{c}$ mass. (c, d) Simultaneous fit to the $M(\gamma\pi^{+}\pi^{-}\eta)$ and $M(\gamma\gamma\pi^{+}\pi^{-})$ distributions.
}
\label{fig:scatter}
\end{center}
\end{figure}

After the above requirements, the momenta of final states are updated according to the kinematic fit for the further analysis.
The distributions of the invariant mass $M(\gamma\pi^{+}\pi^{-}\eta)$ or $M(\gamma\gamma\pi^{+}\pi^{-})$ versus the  invariant mass $M(\pi^{+}\pi^{-}\eta)$ or $M(\gamma\pi^{+}\pi^{-})$) of candidate events in data surviving the selection criteria are shown in Figs.~\ref{fig:scatter} (a, b). Throughout this paper, $M$ denotes the invariant mass of an individual particle combination. Clear $h_c$ signals are observed in the $\gamma\eta^\prime$ final states. The $\eta^\prime$ signal region is defined as [0.945, 0.970]~GeV/$c^2$ in $M(\pi^{+}\pi^{-}\eta)$ and $M(\gamma\pi^{+}\pi^{-})$, while its sideband regions are defined as [0.900, 0.925] and [0.990, 1.015]~GeV/$c^{2}$ for the two $\eta^\prime$ decay modes. Figures~\ref{fig:scatter} (c, d) show the one-dimensional projections of $M({\gamma\pp\eta})$ and $M(\GG\pp)$ for candidate events in the $\eta^\prime$ signal region.  Events from the sideband regions are superimposed, normalized by a 0.5 scale factor.  No obvious peaking background is found in the $\eta^\prime$ sideband regions. Therefore, the background is described by an ARGUS function~\cite{argus}, with the threshold parameter fixed to the kinematic threshold of 3.551 ${\rm GeV}/c^2$, while the other parameters are allowed to float. 
The signal shape is modeled using a combination of Gaussian kernels, implemented using the Roofit RooHistPdf~\cite{Verkerke:2003ir}

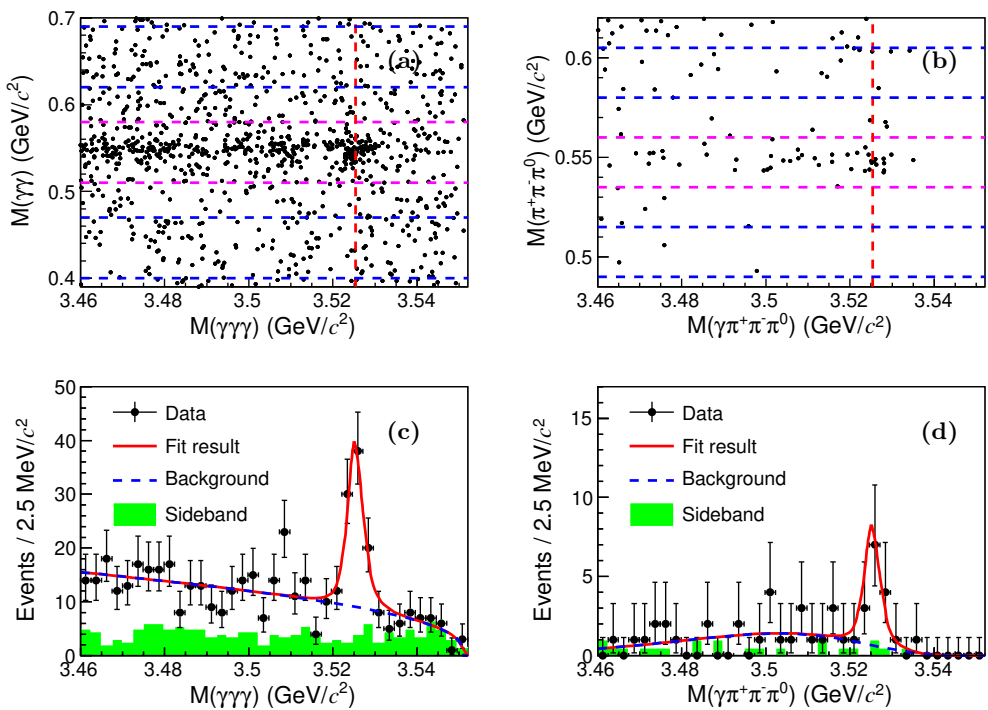
\begin{figure}[htbp]
\begin{center}
\hspace*{-0.3cm}
\begin{overpic}[width=0.9\textwidth]{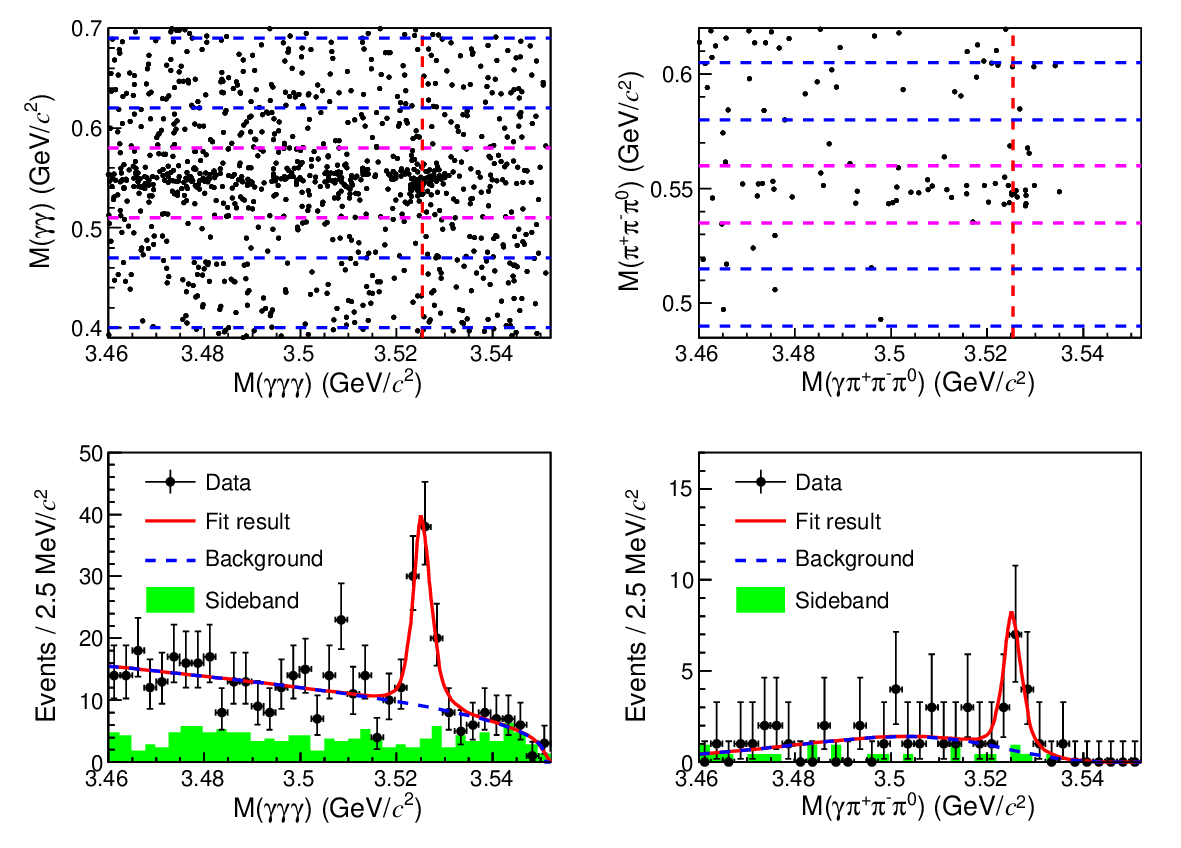}
\put(150,250){\footnotesize{\textbf{(a)}}}
\put(350,250){\footnotesize{\textbf{(b)}}}
\put(150,110){\footnotesize{\textbf{(c)}}}
\put(350,110){\footnotesize{\textbf{(d)}}}
\end{overpic}
\caption{Distributions of (a) $M(\gamma\gamma\gamma)$ versus $M(\gamma\gamma)$  of the candidates for $h_{c}\too\gamma\eta$ with $\eta\too\gamma\gamma$, and (b) $M(\gamma\pi^{+}\pi^{-}\piz)$ versus $M(\pi^{+}\pi^{-}\piz)$ of the candidates for $h_{c}\too\gamma\eta$ with $\eta\too\pi^{+}\pi^{-}\piz$. The horizontal pink (blue) dashed lines mark the signal (sideband) region(s) of $\eta$ and the vertical red dashed lines mark the known $h_{c}$ mass. (c, d) Simultaneous fit to the $M(\gamma\gamma\gamma)$ and $M(\gamma\pi^{+}\pi^{-}\piz)$ distributions.
}
\label{fig:fit_sim}
\end{center}
\end{figure}

\begin{figure}
\begin{center}
\hspace*{-0.5cm}
\begin{overpic}[width=0.45\textwidth]{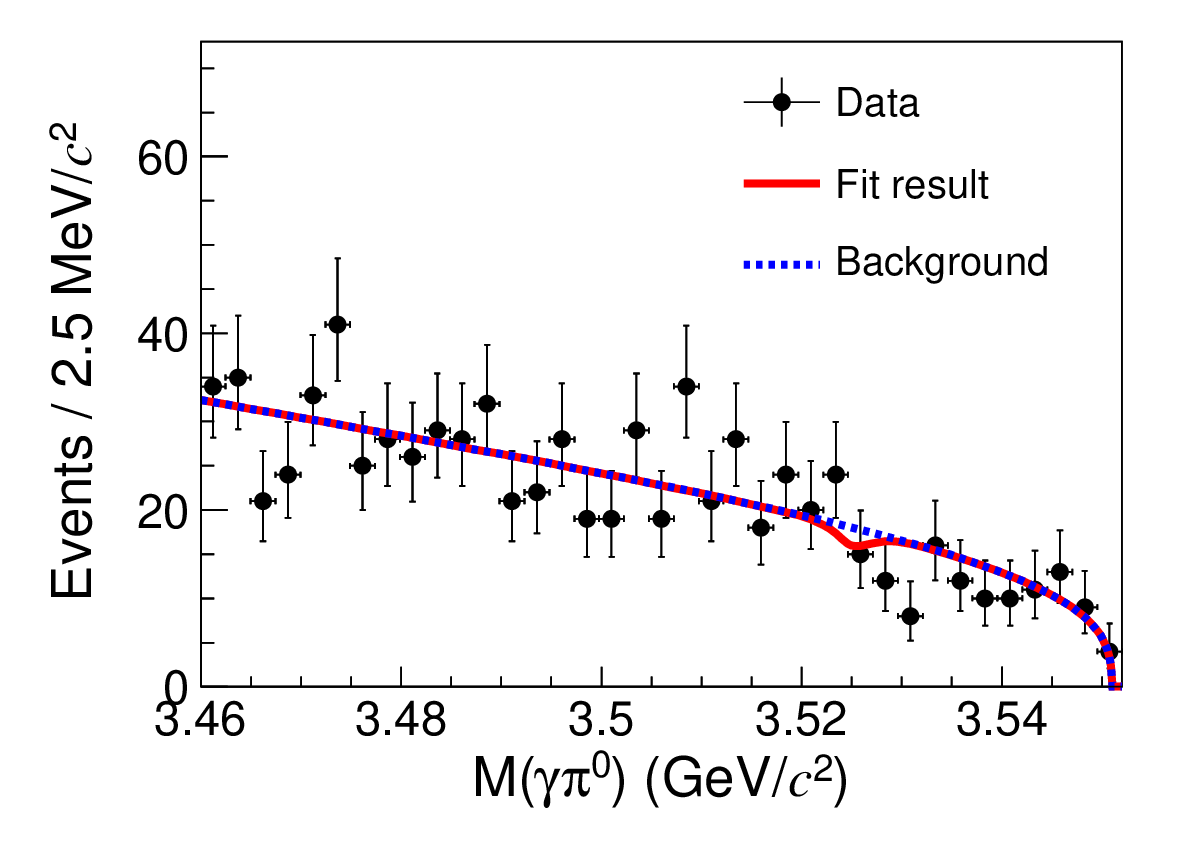}
\end{overpic}
\begin{overpic}[width=0.45\textwidth]{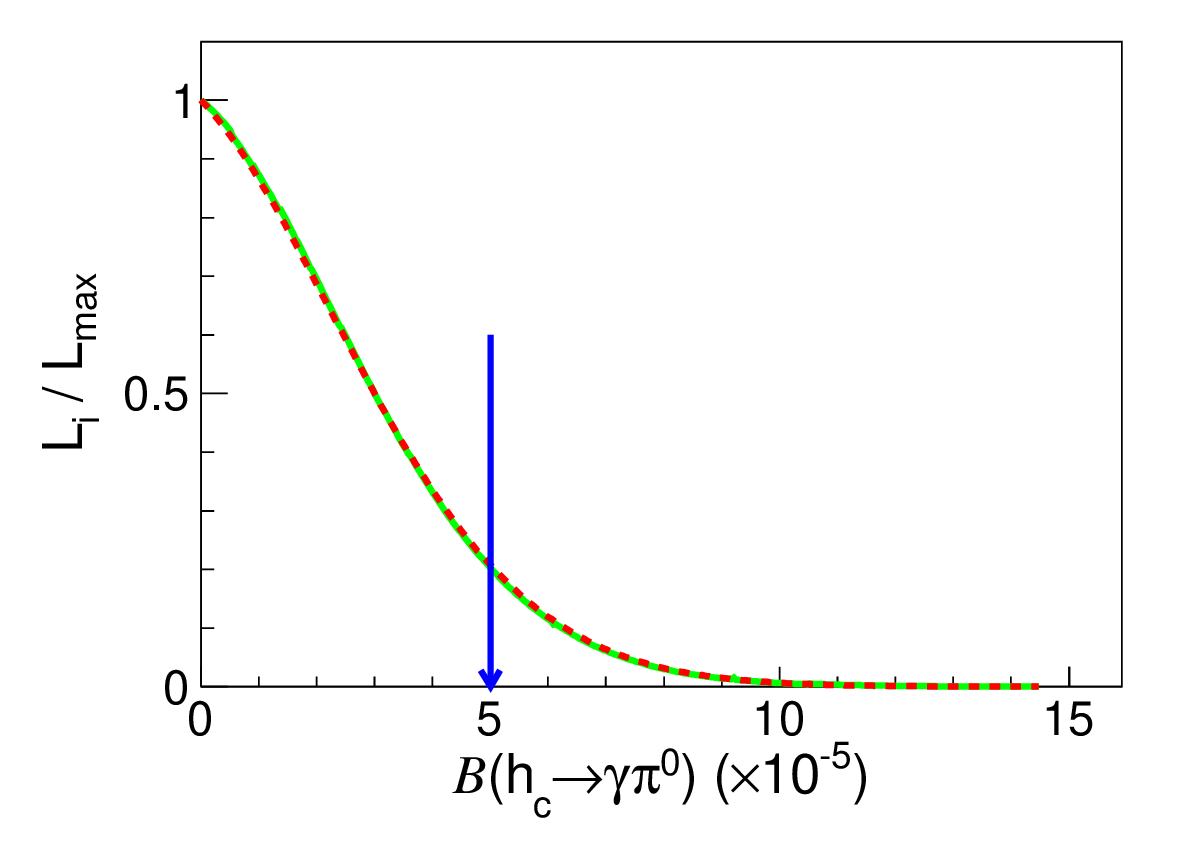}
\end{overpic}
\hspace*{-0.5cm}
\caption{(left) Fit to the $M(\gamma\piz)$ distribution of  $h_c\too\gamma\piz$ candidates.
 (right) The green solid line represents the normalized likelihood distribution incorporating additive systematic uncertainties, while the red dashed line corresponds to the aforementioned green solid line with the inclusion of multiplicative systematic uncertainties. The blue arrow indicates the upper limit on the branching fraction at the 90\% C.L.}
\label{fig:fit_gampi0}
\end{center}
\end{figure}

Similarly, the distributions of $M(\gamma\gamma\gamma)$ or $M(\gamma\pi^{+}\pi^{-}\piz)$ versus $M(\gamma\gamma)$ or $M(\pi^{+}\pi^{-}\piz)$ of candidate events in data are shown in Figs.~\ref{fig:fit_sim}(a, b). Significant $\eta$ and $h_{c}$ signals are observed. For the $\eta\too\gamma\gamma$ mode, the $\eta$ signal region lies within  $[0.510,~0.580]~ {\rm GeV}/c^{2}$, with its sideband regions defined to be [0.400, 0.470] and $[0.620,~0.690]~{\rm GeV}/c^{2}$. For the $\eta\too\pi^{+}\pi^{-}\piz$ mode, the $\eta$ signal region is defined as $[0.535,~0.560]~{\rm GeV}/c^{2}$, with its $\eta$ sideband regions defined as [0.490, 0.515] and $[0.580,~0.605]~{\rm GeV}/c^{2}$. Figures~\ref{fig:fit_sim} (c, d) present the one-dimensional projections of $M(\gamma\GG)$ and $M(\gamma\ppp)$ for candidate events in the $\eta$ signal region, with background events from the sideband region superimposed, again normalized by a scale factor of 0.5. As there is no significant peaking background evident in the $\eta$ sideband regions, we utilize similar signal and background functions to those employed in the $h_c\too\gamma\eta^\prime$ fits. Figure~\ref{fig:fit_gampi0} (left) shows the $M(\gamma\piz)$ from data for the process $h_{c}\too\gamma\pi^{0}$, and no significant signal is observed.

The branching fraction of $h_c\too\gamma P (P=\eta^\prime,~\eta,~\piz)$ is calculated by
\begin{linenomath*}
\begin{equation}
\begin{aligned}
    \mathscr{B}(h_c\too\gamma P) =\frac{N^{\rm sig}}{N^{\rm{tot}}\cdot{\mathscr{B}(\psi(3686)\too \pi^{0}h_c)\prod_{i}\mathscr{B}_{i}\epsilon}},
\end{aligned}
\end{equation}
\end{linenomath*}
where $N^{\rm sig}$ is the signal yield, $\prod_{i}\mathscr{B}_{i}$ is the product of the branching fractions of the intermediate decays in the decay chain of interest from the PDG~\cite{pdg},
$\epsilon$ is the detection efficiency, and $N^{\rm{tot}}$ is the total number of $\psi(3686)$ events.

For the decay of $h_c\too\gamma\eta^\prime$, two final states are used to reconstruct the $\eta^\prime$ meson. A simultaneous unbinned maximum-likelihood fit is performed to determine the branching fraction $\mathscr{B}(h_c\too\gamma\eta^\prime)$, which is taken as a common fit parameter for both $\eta^\prime$ decay modes. The number of $h_c$ signal events in the two different final states is calculated by
\begin{linenomath*}
\begin{equation}
\begin{aligned}
&N^{\rm sig}_{X}=N^{\rm{tot}}\cdot{\mathscr{B}(\psi(3686)\too \pi^{0}h_c)}\cdot{\mathscr{B}(\pi^{0}\too\gamma\gamma)}\cdot\mathscr{B}(h_c\too\gamma\eta^\prime)\cdot{\mathscr{B}(\eta^\prime\too X)}\cdot\epsilon_{X},
\end{aligned}
\end{equation}
\end{linenomath*}
where $X$ denotes one of the final states of the $\eta^\prime$ decay. The projections on $M({\gamma\pp\eta})$ and $M(\GG\pp)$ of the simultaneous fit are presented in Figs.~\ref{fig:scatter} (c, d).

Similarly, a simultaneous fit is used to extract the branching fraction of $h_{c}\too\gamma\eta$, which is treated as a common fit parameter for both $\eta$ decay modes. The projections on $M(\gamma\GG)$ and $M(\gamma\ppp)$ of  the simultaneous fit are shown in Figs.~\ref{fig:fit_sim} (c, d). The $h_c\too\gamma\eta$ signal significance is estimated to be $9.0\,\sigma$, which is determined by the change of the log-likelihood value and the number of degrees of freedom in the fit with and without the $h_{c}$ signal. The measured branching fractions  for $h_c\too\gamma\eta^\prime$ and  $h_c\too\gamma\eta$, and their ratio, are listed in TABLE~\ref{tab:result}.

\begin{table*}[htbp]
\centering
\scriptsize
\begin{tabular}{cccccccccc}
\hline
            &    \ \  \ \ $\mathscr{B}(h_c\too\gamma\eta^\prime)$($\times10^{-3}$)  \ \ \ \ \ \ \ \  &  \ \ \ \ \  $\mathscr{B}(h_c\too\gamma\eta)$($\times10^{-4}$)  \ \ \ \ \  &   \ \ \ \ \ $\mathscr{B}(h_c\too\gamma\eta)/\mathscr{B}(h_c\too\gamma\eta^\prime)$ (\%) \\
\hline
This work &  $1.40\pm0.11\pm0.04\pm0.10$        & $3.77\pm0.55\pm0.13\pm0.26$  & $27.0\pm4.4\pm1.0$ \\
BESIII~\cite{hcgamX}      &  $1.52\pm0.27\pm0.29$               &  $4.7\pm1.5\pm1.4$           & $30.7\pm11.3\pm8.7$ \\
\hline
\end{tabular}
\caption{Measured branching fractions for $h_c\too\gamma\eta^\prime$ and $h_c\too\gamma\eta$, and the ratio of branching fractions for the two decay modes. The first uncertainties are statistical, the second systematic, and the third from the branching fraction of $\psi(3686)\too\pi^{0}h_c$, which cancels and so does not appear in the ratio. Also shown are the previous results from BESIII~\cite{hcgamX}.}
\label{tab:result}
\end{table*}

No significant signal is observed for $h_c\too\gamma\piz$, as can be seen in the left panel of Fig.~\ref{fig:fit_gampi0}. An upper limit of $5.0\times10^{-5}$ is set on the branching fraction of $h_c\too\gamma\piz$ at the 90\% confidence level (C.L.), using a profile likelihood fit, as shown in the right panel of Fig.~\ref{fig:fit_gampi0}. The green solid line represents the likelihood distribution incorporating additive systematic uncertainties, while the red dashed line is the convolution of the likelihood curve (green solid line) with multiplicative systematic uncertainties following the method described in Ref~\cite{para}. Detailed descriptions of the systematic uncertainties are provided in the subsequent section.

\section{Systematic uncertainties}

The systematic uncertainties in the measurements of $\mathscr{B}(h_c\too\gamma\eta^\prime(\eta, ~\piz))$ arise from the understanding of the pion tracking, the photon detection, the $\pi^{0}$ reconstruction, the $\eta$ mass window, the $R$ ratio requirement for $h_c\too\gamma\eta$, $\eta\too\gamma\gamma$, the $\cos\theta_{\pi^{0}_{1}\pi^{0}_{2}}$ requirement for the $h_c\too\gamma\piz$ selection, the kinematic fit, the signal shape, the background shape, the assumed  branching fractions, and the number of $\psi(3686)$ events, as listed in TABLE~\ref{tab:summererror}.

\begin{table*}[htbp]
\centering
\scriptsize
\begin{tabular}{c|cc|cc|cccc}
\hline
Source    &   \ \   $\eta^\prime\too\pi^{+}\pi^{-}\eta$   \ \  &     \ \    $\eta^\prime\too\pi^{+}\pi^{-}\gamma$  \ \   &   $\eta\too\gamma\gamma$ \ \  &      \ \  $\eta\too\pi^{+}\pi^{-}\pi^{0}$  \ \   &   \ \     $\piz\too\gamma\gamma$  \\
\hline
Tracking                     &  2.0  &  2.0  &  - &  2.0  &   -  \\
Photon detection             & 1.5   &  1.0  & 1.5  &  0.5  &  0.5  \\
$\pi^{0}$  reconstruction    & 0.5   & 0.5   & 0.5  & 2.4   & 6.4 \\
$\eta$ mass window           & 1.0   & -  & 1.0   & -   & -  \\
$R_{E}$ ratio and $\cos\theta_{\pi^{0}_{1}\pi^{0}_{2}}$        & -   & -   & 0.8  & -   & -  \\
Kinematic fit                & 0.9   &  0.6  & 1.5  &  1.5  & 2.8 \\
Signal shape         &  \multicolumn{2}{c|}{0.1}  &  \multicolumn{2}{c|}{0.1} & - \\
Background shape     &  \multicolumn{2}{c|}{0.3}  &  \multicolumn{2}{c|}{2.0} & - \\
$\mathscr{B}_{\eta^\prime,~\eta,~\piz}$ & \multicolumn{2}{c|}{1.0} & \multicolumn{2}{c|}{0.4}  & 0.03 \\
$\mathscr{B}(\psi(3686)\too\pi^{0}h_c)$  & \multicolumn{2}{c|}{6.8} & \multicolumn{2}{c|}{6.8}  & 6.8 \\
Number of $\psi(3686)$ events    & \multicolumn{2}{c|}{0.5}  & \multicolumn{2}{c|}{0.5}  & 0.5 \\
Sum            &   \multicolumn{2}{c|}{7.3} &  \multicolumn{2}{c|}{7.5}  & 9.8 \\
\hline
\end{tabular}
\caption{Relative systematic uncertainties (\%) on the branching-fraction measurements, categorised by the decay chain used to reconstruct the final state.  A dash (-) indicates that the source is not relevant for that decay.}
\label{tab:summererror}
\end{table*}

\begin{itemize}
\item \emph{Tracking.} The uncertainty arising from the tracking efficiency is assigned to be $1.0\%$ for each pion from studies of a  control sample of $\psi(3686)\too\pi^{+}\pi^{-}J/\psi$, $J/\psi\too l^{+}l^{-}$ decays~\cite{track}.

\item \emph{Photon detection.} The uncertainty in the photon-detection efficiency is studied using a control sample of the $\EE\too\gamma\mu^{+}\mu^{-}$ events. The four-momentum of the initial-state-radiation photon is predicted by the $\mu^{+}\mu^{-}$ pair. The photon detection efficiency is defined as the fraction of reconstructed photons with four-momentum matching in the EMC. The systematic uncertainty, defined as the relative difference in efficiency between data and MC simulation, due to photon reconstruction efficiency is observed always less or equal to 0.5\%. Therefore, the conservative systematic uncertainty of 0.5\% is assigned to each photon from photon detection efficiency.

\item \emph{$\pi^{0}$ reconstruction and $\eta$ mass window.} The uncertainty of the $\pi^{0}$ reconstruction efficiency is estimated after taking account of its momentum distribution. The relative difference of the $\piz$ reconstruction efficiencies between data and MC simulation is obtained from a study of  $\EE\too\omega\piz$ events at $\sqrt{s}=3.773$ GeV. This relative difference is found to decrease linearly, ($0.06-2.41\times p$)\%, as a function of momentum $p$. The associated systematic uncertainties for each decay is listed in TABLE~\ref{tab:summererror}. The uncertainty from the $\eta$ mass window for the $\eta\too\gamma\gamma$ mode is determined to be 1\% per $\eta$ through the study of a high-purity control sample of $J/\psi\too p\bar{p}\eta$ events~\cite{track}.

\item \emph{Kinematic fit.} The uncertainty associated with the kinematic fit arises from the charged-track helix parameter correction and the difference in the photon-energy resolution between data and MC simulation. For charged tracks, a correction is made to account for the difference in the distribution of track-helix parameters that is observed between data and MC in a  control sample of $J/\psi\too K^+K^-\pi^+\pi^{-}$ events~\cite{kinematic}. The resolution of the reconstructed photon energy in MC is degraded by 4\% to match  what is observed in data. The difference between the efficiencies with and without these corrections is taken as the systematic uncertainty.

\item \emph{$R_{E}$ ratio and $\cos\theta_{\pi^{0}_{1}\pi^{0}_{2}}$.} The systematic uncertainty associated with the $R_{E}$ ratio ($R_{E} ={|E_{\gamma1}-E_{\gamma2}|}/{P_{\eta}}$) is estimated by an alternative analysis without requiring $R_{E}<0.94$, and the difference in the final result is taken as the systematic uncertainty. The uncertainty arising from the $\cos\theta_{\pi^{0}_{1}\pi^{0}_{2}}>-0.5$ requirement is obtained by varying the cut by $\pm0.1$, and taking the maximum deviation observed from the baseline result as the corresponding systematic uncertainty.

\item \emph{Signal shape.} The uncertainty associated with the signal shape is estimated by convolving the MC-determined shape with a Gaussian function characterized by free parameters. For the decay of $h_c\too\gamma\eta$, two Gaussian functions from the signal shapes of $\eta\too\gamma\gamma$ and $\eta\too\pi^{+}\pi^{-}\pi^{0}$ share the same parameters due to the low sample size for the $\eta\too\pi^{+}\pi^{-}\pi^{0}$ mode. For the $h_{c}\too\gamma\piz$ decay, due to no signal being observed, the Gaussian parameters are fixed to those from the fit to the  $h_c\too\gamma\eta,~\eta\too\gamma\gamma$ mass distribution. The difference in the final results is taken as the systematic uncertainty.

\item \emph{Background shape.} To assess the uncertainty associated with the description of the background, its shape in the fit is changed from that of the default analysis to be a second-order polynomial function. The difference between the two fit results is assigned as the systematic uncertainty.

\item \emph{Input branching fractions.} The uncertainties of the branching fractions are taken from the results reported in  the PDG~\cite{pdg}.

\item \emph{Number of $\psi(3686)$ events.} The total number of $\psi(3686)$ events is determined by using  inclusive hadronic decays, as described in Ref.~\cite{0912data}. The uncertainty on the number of $\psi(3686)$ events is estimated to be 0.5\%.

\end{itemize}

The systematic uncertainty on the upper limit of signal events $N^{\rm up}_{h_c\too\gamma\piz}$ at 90\% C.L. includes contributions from both additive and multiplicative sources. The additive sources comprise those from the signal shape, background shape, and $\cos\theta_{\pi^{0}_{1}\pi^{0}_{2}}>-0.5$ requirement. Each of them is considered separately, and the upper limit is recomputed. The largest value, which comes from the variation in signal shape, is chosen as the conservative estimation. The other sources listed in TABLE~\ref{tab:summererror} are multiplicative. To include multiplicative systematics,
 the obtained likelihood curve is used to convolute a Gaussian function in which the uncertainty is taken as a parameter~\cite{para}. The likelihood curves before and after consider the uncertainty are shown in the right panel of Fig.~\ref{fig:fit_gampi0}.

The relative systematic uncertainties in the branching-fraction measurements are summarized in TABLE~\ref{tab:summererror}. The systematic uncertainties depend on the $\eta$ or $\eta^\prime$ decay modes, in which the uncertainty from the tracking, the photon detection, the $\pi^{0}$ reconstruction, the $\eta$ mass window, $R$ ratio, and kinematic fit, are combined by a weighted average, considering the number of signal events for each case~\cite{sysmethod}. The total systematic uncertainty is obtained by summing all contributions in quadrature, assuming they are independent of each other.

\section{Summary}

We perform the studies of the decays of $h_c\to\gamma\eta^\prime$, $\gamma\eta$ and $\gamma\piz$ via $\psi(3686)\too\pi^{0}h_{c}$, using $(27.12\pm0.14)\times10^{8}$ $\psi(3686)$ events collected by the BESIII detector.
The decay $h_c\too\gamma\eta$ is observed for the first time with a statistical significance of $9.0\,\sigma$ and its branching fraction is measured to be $\mathscr{B}(h_c\too\gamma\eta)=(3.77\pm0.55\pm0.13\pm0.26)\times10^{-4}$, where the first uncertainty is statistical, the second systematic, and the third arises from the knowledge of the branching fraction of $\psi(3686)\too\pi^{0}h_c$.
Additionally, an improved measurement of $\mathscr{B}(h_c\too\gamma\eta^\prime)=(1.40\pm0.11\pm0.04\pm0.10)\times10^{-3}$ is obtained. From these results, the ratio $R_{h_c}=\frac{\mathscr{B}(h_c\too\gamma\eta)}{\mathscr{B}(h_c\too\gamma\eta^\prime)}$ is calculated to be $(27.0\pm4.4\pm1.0)\%$, where the common systematic uncertainties between the two branching fractions are accounted for. These measurement results are consistent with the previous measurements~\cite{hcgamX}, with the improved precision by more than a factor of two. Our result favors the prediction in Ref.~\cite{mixpqcd}, suggesting comparable contributions from both the quark-antiquark content and the gluonic content in $\eta^{(\prime)}$. Furthermore, these improved measurements offer valuable prospects for the study of SU(3)-flavor symmetries~\cite{u3f}. We firstly search for $h_{c}\too\gamma\pi^{0}$, no significant signal is found, and an upper limit of $\mathscr{B}(h_c\too\gamma\pi^{0})<5.0\times10^{-5}$ is set on its branching fraction at the 90\% C.L.

\acknowledgments
The BESIII Collaboration thanks the staff of BEPCII and the IHEP computing center for their strong support. This work is supported in part by National Key R\&D Program of China under Contracts Nos. 2020YFA0406300, 2020YFA0406400, 2023YFA1606000; National Natural Science Foundation of China (NSFC) under Contracts Nos. 12375070, 12375071, 11635010, 11735014, 11935015, 11935016, 11935018, 12025502, 12035009, 12035013, 12061131003, 12192260, 12192261, 12192262, 12192263, 12192264, 12192265, 12221005, 12225509, 12235017, 12361141819; the Chinese Academy of Sciences (CAS) Large-Scale Scientific Facility Program; the CAS Center for Excellence in Particle Physics (CCEPP); Joint Large-Scale Scientific Facility Funds of the NSFC and CAS under Contract No. U2032108, U1832207; 100 Talents Program of CAS; The Institute of Nuclear and Particle Physics (INPAC) and Shanghai Key Laboratory for Particle Physics and Cosmology; German Research Foundation DFG under Contracts Nos. 455635585, FOR5327, GRK 2149; Istituto Nazionale di Fisica Nucleare, Italy; Knut and Alice Wallenberg Foundation under Contracts Nos. 2021.0174, 2021.0299; Ministry of Development of Turkey under Contract No. DPT2006K-120470; National Research Foundation of Korea under Contract No. NRF-2022R1A2C1092335; National Science and Technology fund of Mongolia; National Science Research and Innovation Fund (NSRF) via the Program Management Unit for Human Resources \& Institutional Development, Research and Innovation of Thailand under Contract No. B16F640076; Polish National Science Centre under Contract No. 2019/35/O/ST2/02907; Swedish Research Council under Contract No. 2019.04595; The Swedish Foundation for International Cooperation in Research and Higher Education under Contract No. CH2018-7756; U. S. Department of Energy under Contract No. DE-FG02-05ER41374.

\clearpage

\section*{The BESIII Collaboration}
\addcontentsline{toc}{section}{The BESIII Collaboration}
\begin{small}
M.~Ablikim$^{1}$, M.~N.~Achasov$^{4,c}$, P.~Adlarson$^{75}$, O.~Afedulidis$^{3}$, X.~C.~Ai$^{80}$, R.~Aliberti$^{35}$, A.~Amoroso$^{74A,74C}$, Q.~An$^{71,58,a}$, Y.~Bai$^{57}$, O.~Bakina$^{36}$
% ...
\\
 {\it
$^{1}$ Institute of High Energy Physics, Beijing 100049, People's Republic of China\\
% ...
$^{3}$ Bochum Ruhr-University, D-44780 Bochum, Germany\\
$^{4}$ Budker Institute of Nuclear Physics SB RAS (BINP), Novosibirsk 630090, Russia\\
% ...
$^{35}$ Johannes Gutenberg University of Mainz, Johann-Joachim-Becher-Weg 45, D-55099 Mainz, Germany\\
$^{36}$ Joint Institute for Nuclear Research, 141980 Dubna, Moscow region, Russia\\
% ...
$^{57}$ Southeast University, Nanjing 211100, People's Republic of China\\
$^{58}$ State Key Laboratory of Particle Detection and Electronics, Beijing 100049, Hefei 230026, People's Republic of China\\
% ...
$^{71}$ University of Science and Technology of China, Hefei 230026, People's Republic of China\\
% ...
$^{74}$ University of Turin and INFN, (A)University of Turin, I-10125, Turin, Italy; (B)University of Eastern Piedmont, I-15121, Alessandria, Italy; (C)INFN, I-10125, Turin, Italy\\
$^{75}$ Uppsala University, Box 516, SE-75120 Uppsala, Sweden\\
% ...
$^{80}$ Zhengzhou University, Zhengzhou 450001, People's Republic of China\\
\vspace{0.2cm}
$^{a}$ Deceased\\
% ...
$^{c}$ Also at the Novosibirsk State University, Novosibirsk, 630090, Russia\\
% ...
}

\vspace{0.4cm}
\end{small}

\end{document}